\documentclass[10pt,aps,prl,groupedaddress,twocolumn,amsmath,amssymb,showpacs]{revtex4-1}
\usepackage[latin1]{inputenc}
\usepackage{amsmath}
\usepackage{amsfonts}
\usepackage{amssymb}
\usepackage{natbib}
\usepackage[pdftex]{graphicx}
\usepackage{dcolumn}
\usepackage{bm}
\usepackage{epstopdf}
\begin{document}

\title{Fractional Quantum Hall Physics in  Jaynes-Cummings-Hubbard Lattices}
\author{Andrew L.C. Hayward$^1$}
\author{Andrew M. Martin$^1$}
\author{Andrew D. Greentree$^{1,2}$}
\affiliation{$^1$School of Physics, University of Melbourne, Vic 3050, Australia}
\affiliation{$^2$Applied Physics, School of Applied Sciences, RMIT University, Vic 3001, Australia}

\date{\today}
        
\begin{abstract}
Jaynes-Cummings-Hubbard arrays provide unique opportunities for quantum
 emulation as they exhibit convenient state preparation and measurement, and
in-situ tuning of parameters.
 We show how to realise strongly correlated states of light
in Jaynes-Cummings-Hubbard arrays under the introduction of an
effective magnetic field. The effective field is realised by dynamic tuning of
the cavity resonances.
  We demonstrate the existence of Fractional Quantum Hall states by computing
  topological invariants, phase transitions between topologically distinct
  states, and Laughlin wavefunction overlap.         
 
\end{abstract}

\maketitle

Quantum systems with highly correlated states exhibit an
exponential growth of Hilbert space with the number of particles, making the
study of arbitrary states of even modest systems computationally intractable.
This problem has motivated efforts in the field of quantum
emulation\cite{Buluta2009}.  A quantum emulator is designed to replicate the
physics of some target system.  Such emulators require scalable and convenient
state preparation and measurement, and control over single and many-body
interactions. Proposals for emulation platforms include ultra cold-atoms,
superconductors, and
superfluids\cite{Lewenstein2007,Oudenaarden1996,Mostame2008}.   Another interesting
platform is coupled atom-cavity
systems\cite{Hartmann2006,Greentree2006,Angelakis2007,Na2008,Lei2008} and
there has been significant progress towards realising this
goal\cite{Birnbaum2005,Houck2007, Notomi2008,Lepert2011}.  Here we explore the
physics of the Fractional Quantum Hall Effect (FQHE) as it relates to atom-cavity
   systems. 

Thirty years after their discovery, the Integer\cite{Klitzing1980} and
Fractional\cite{Tsui1982} Quantum Hall Effects are still the focus of intense
theoretical and experimental attention\cite{Bolotin2009,Storni2010}. The FQHE relies on the presence of particle-particle interactions to form
highly correlated states. These states can exhibit anyonic, and sometimes
non-abelian, excitations which are explicitly non-local. As such, the
investigation of large systems suffers strongly from the exponential explosion
in Hilbert space. While there exist exact solutions for some FQHE
systems, such as the Laughlin ansatz\cite{Laughlin83}, these have yet to be observed directly
in experiment. For this reason, emulation of the FQHE,
particularly emulating the strong magnetic fields required, has become a major
topic of interest in the scientific
community\cite{Jaksch2003,Sorensen2005,Kolovsky2011a,Fetter2010}.
\begin{figure}
 \label{fig:nicepic}
   \begin{center} \includegraphics[width = 8cm, height =
   8cm]{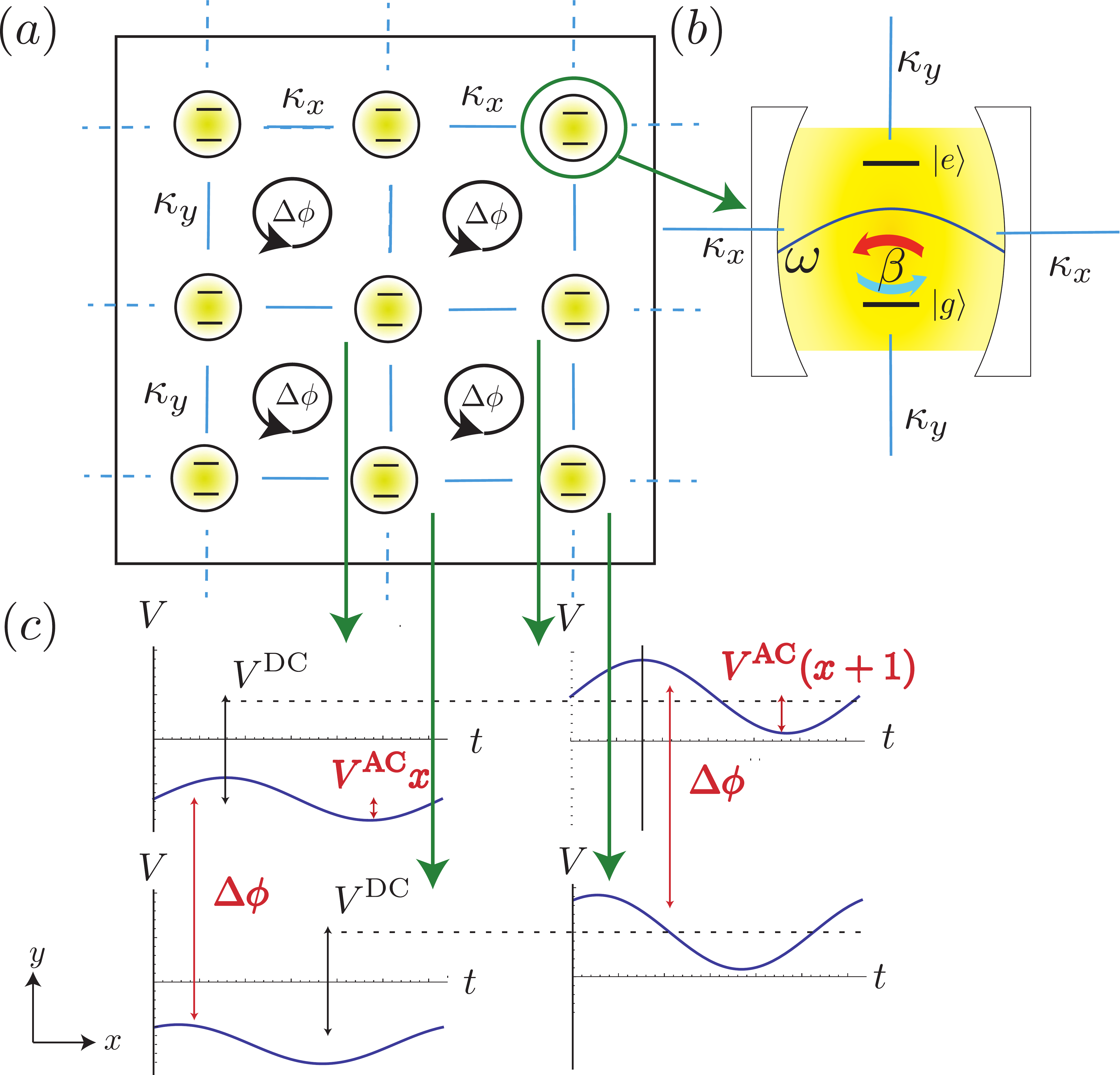} \end{center} \caption{(a) Schematic of a square JCH
   lattice with a constant effective magnetic field. Photons moving around a plaquette
   acquire a  phase $\Delta\phi$. (b) A single mode  photonic cavity
   with frequency $\omega$ coupled to a two level atom with strength
   $\beta$. (c) Scheme for breaking TRS in photonic cavities: a potential $V =
   \left[V^{\mathrm{DC}} + V^{\mathrm{AC}}\sin{(\omega^{\mathrm{rf}}t +
   \Delta\phi{}y)}\right]x$ ($x$ and $y$ in units of the lattice spacing) is
   applied to the cavities (indicated by green
   arrows) by dynamically tuning $\omega$. The phase offset, $\Delta\phi$, along $y$ results in the synthetic magnetic
   field seen in (a).}\vspace{-.5cm}
\end{figure}

Here we show the existence of FQHE states in the
   Jaynes-Cummings-Hubbard (JCH)
   model\cite{Hartmann2006,Greentree2006,Angelakis2007} in the presence of an
   artificial magnetic field. These states constitute new, strongly correlated
   states of light. A JCH lattice consists of an array of coupled
   photonic cavities, with each cavity mode coupled to a two-level atom [see
   Figs. 1(a) and (b)].  JCH
   systems promise unparallelled control and readout of the full quantum
   mechanical wavefunction. The JCH model is predicted to exhibit a number of
   solid state phenomena, including Mott/superfluid
   phases\cite{Greentree2006}, semi-conductor physics\cite{Quach2009},
   Josephson effect\cite{Lei10}, metamaterials properties\cite{Quach11},  and
   Bose-glass phases\cite{Rossini2007}.

We begin by introducing the JCH model, and discuss how an artificial magnetic
field can be created in a photonic cavity system. To demonstrate FQHE
physics, we compute the groundstates for small systems. These groundstates are
compared to a modified Laughlin ansatz, and their topology investigated.

Each cavity in the JCH lattice is described by the Jaynes-Cummings (JC)  Hamiltonian
\begin{equation}
   H^{JC} = \omega{}L + \Delta\sigma{}^{+}\sigma^{-} +
   \beta{}(\sigma^{+}a+\sigma^{-}a^{\dagger}) \mathbb{,}
   \label{equ:JC}
\end{equation}
   where $a$ is the photonic annihilation operator, $\sigma^{\pm}$ are the
   atomic raising and lowering operators, $\Delta$ the atom-photon
   detuning, $\beta$ the coupling energy and $\hbar{}=1$. The states
   $|g (e),n\rangle$, where $n$ is the number of photons, and $g(e)$ are the
   ground (excited) state of the atom, form the single cavity basis.
   $H^{JC}$ commutes with the total excitation number operator,
   $L=a^{\dagger}a + \sigma^{+}\sigma^{-}$. Therefore the total excitations in
   the cavity, $\ell$, is a good quantum number.  The eigenstates of
   Eq.~(\ref{equ:JC}) are termed polaritons, superpositions of atomic and photonic excitations,
   and are a function of $\ell$ and $\Delta/\beta$. 
   
   The JCH model describes a tight-binding JC lattice: 
 \begin{equation} H^{JCH} = H^{JC} + K = 
    \sum\limits^{N}_{i}H^{JC}_i  
    -\sum\limits_{<i,j>}\kappa_{ij}a_i^\dagger{}a_j
    \label{equ:JCH}
 \end{equation}
where $\kappa_{ij}$ is the tunneling rate between cavities $i$ and $j$ and the
sum over $\langle{}i,j\rangle$ is between nearest neighbors only.

For large detuning ($|\Delta| \gg \beta$), 
   eigenstates separate out into either atomic or photonic modes.  In this limit,
   the photonic or atomic mode can be adiabatically eliminated.  Eliminating
   the atomic modes, the photonic mode has a weak Kerr-type photon-photon
   repulsion\cite{Na2008}
and the exchange of energy between atomic and photonic modes is strongly
suppressed.
However,
virtual processes lead to effective interactions in the photonic and atomic
submanifolds.  
Photons have an atomic mediated non-linear onsite repulsion,
making  the JCH  model equivalent to the Bose-Hubbard (BH) model
\cite{Hohenadler11}.  Atomic modes are coupled with the effective hopping rate
$\kappa^{eff}_{ij}=\kappa_{ij}\beta^2/\Delta^2$\cite{Makin2009}.  As the
atomic modes are restricted to two levels, this is effectively a  hardcore boson
field for atomic states, in contrast to the weakly-interacting photon field.
\begin{figure}

\includegraphics[scale=.45]{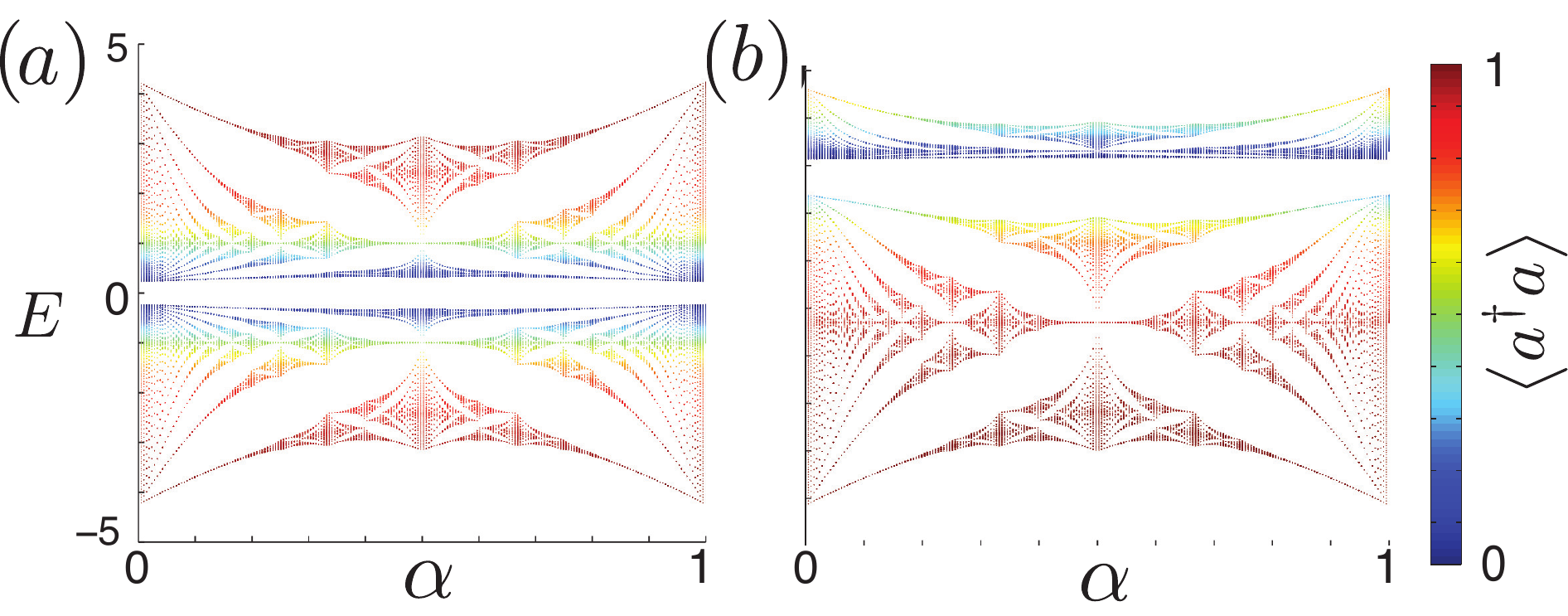}
\caption{\label{fig:singlespec}Single particle spectrum
of the JCH lattice with (a) $\Delta=0$ and (b)  $\Delta=3$. In each
case
$\kappa=1$ and $\beta=1$. The spectra comprise of two transformed
Hofstadter butterflies. Color indicates the projection into the photonic
modes.}\vspace{-0.5cm}
\end{figure}

The QHE occurs in a 2DEG in the presence of a perpendicular magnetic field,
which breaks time reversal symmetry (TRS).  Any mechanism which breaks TRS
manifests in the Hamiltonian as a vector potential. An artificial magnetic
field may then be realized via the introduction of some TRS breaking
interaction.  Constant rotation or linear acceleration are the classic
examples, leading to constant synthetic magnetic and electric fields
respectively. Cho \emph{et al.}\cite{Cho2008} propose a scheme for TRS
breaking in multi-mode cavities with far detuned atoms. Fang {\it et al.}\cite{Fang2011}
use magneto-optical resonators in photonic crystals to break TRS, and Koch
{\it et al.}\cite{Koch2010,Nunnenkamp2011} have recently shown TRS breaking in the context of
circuit QED, via the introduction of a special passive coupling element
between microwave resonator junctions.
Here we adapt the methodology proposed by Kolovsky\cite{Kolovsky2011a} to
cavity QED where photon assisted tunneling is used to break TRS. An
electric field with both dc and ac components is applied across one of the
lattice axes [see Fig 1(c)].  Introducing a phase offset, $\Delta\phi$, in the
ac field along the
other axis results in the desired complex coupling, $2\pi\alpha = \Delta\phi$.
The presence of the two fields also leads to modified strengths of $\kappa,
\Delta,$  
and $\beta$, which can
be tuned appropriately to recover the  Hamiltonian in Eq. (1). The dependence
of these parameters on the field is non-trivial, but follows the prescription
in \cite{Kolovsky2011a}.
Photons do not  respond to electric
fields, however, a gradient in the cavity frequency has an equivalent effect.
Recently, cavities with tunable resonances have been
fabricated\cite{Sandberg2009,Johansson2009}. This is achieved by the inclusion of an intra-cavity Josephson
junction, which changes the cavity boundary conditions, and can be tuned via a
magnetic field. Transmission line resonator experiments\cite{Sandberg2009} have shown
$\omega^{\mathrm{rf}}$ can be driven at $\mathcal{O}(10^3)$ times the cavity
dissipation frequency. This ratio provides sufficient time to observe FQH physics.

A magnetic field is defined by a vector potential $\mathbf{A}(x)$, and is
introduced into the Hamiltonian via the minimal substitution,
$\mathbf{p}\rightarrow \mathbf{p}
- q\mathbf{A}(x)$.  On a tight-binding lattice, a vector
potential $\mathbf{A}$ gives rise to a complex hopping rate $\kappa_{ij}
\rightarrow \kappa_{ij}e^{i2\pi\theta_{ij}}$, where $2\pi\theta_{ij} =
\int_{r_i}^{r_j}\mathbf{A}(r)\cdot{}dr$.  As in the continuum case, gauge
symmetry implies that the only physically important parameter is the total
phase, $2\pi\alpha$, picked up around a closed loop, where $\alpha =
\Phi/\Phi_0$ is the fraction of flux quanta through the loop. A constant
magnetic field in the $z$ direction corresponds to a constant $\alpha$ for all
plaquettes on the lattice. Factors of $2\pi$ in the  phase around a loop  are
physically inconsequential, so  we only need consider $\alpha \in [0,1]$.
In a lattice, the effect of larger magnetic field strengths
saturates, as
distinct from the continuum case, where the cyclotron frequency, proportional
to the magnetic field, has no upper bound.

Ignoring the JC term in Eq. (1), the spectrum for the kinetic
term, $K$,  is given by solutions to Harper's equation, resulting
in the famous Hofstadter Butterfly\cite{Hofstadter1976}, a fractal structure,
which has $q$
bands at $\alpha = p/q$, $(p,q) \in \mathbb{Z}$.
For a single excitation on the JCH lattice, the Shr\"{o}dinger Equation is
\begin{equation}
   \label{equ:spjch} \sum\limits_{i}E\psi_{i} =
   \sum\limits_i^{N}[H^{JC}\psi_i  - \kappa{}K_{ij}(\alpha)\psi_j].
\end{equation}
Substituting an eigenvector
$\psi^{K}_i$ of $K$ with energy $k_i$ into
Eq. (\ref{equ:spjch}) yields a single site Hamiltonian: \begin{equation*}
   H^{JC}(k) =  H_0^{JC} + k_i\kappa{}a^\dagger{}a,
\end{equation*}
which is transformed to $H_0^{JC}$ by $\Delta \rightarrow \Delta + k_i$. Thus
the energies and eigenstates are: \begin{equation*} \begin{array}{ll}
   E_{i,\pm} & =-(\Delta+E^H_i)/2 \pm \sqrt{\beta^2  + (\Delta - E^H_i)^2/4}\\
   \psi_{i,\pm}&= \psi^K_i \otimes \psi^{JC}_{\pm}(\Delta -
   k_i\kappa{})\mathbb{.} \end{array} \end{equation*}
   When the detuning is
   small ($\Delta \approx 0$), there are two squashed Hofstadter Butterflies
   with a gap [Fig. 2(a)].  As the relative JC interaction strength $\beta$
   decreases, the two parts converge to recover the original butterfly. 

As previously  discussed, a large detuning separates out the atomic and
photonic states, as shown in Fig. 2. The spectrum is a butterfly with width $\pm 4\kappa$
centered around $-\beta^2/\Delta$, corresponding to the photonic part, and one
with width $\pm 4\kappa^2/\Delta$ centered around $\Delta + \beta^2/\Delta$
[Fig. 2(b)].

The FQHE occurs for systems at sufficiently low temperatures where the flux
filling factor, $\nu={N_p}/{N_\phi}$ is some non-integral rational
$\nu={p}/{q}$, with $N_p$ excitations and $N_\phi$ total flux quanta. Here, particles lie
predominately in the lowest Landau level(LLL). When there is an inter-particle interaction the
groundstate has long range off-diagonal order and an energy gap.  

We study the FQHE on a JCH lattice with periodic boundary conditions, to
avoid edge effects, and 
 focus on the $\nu = 1/2$ state, which is the most stable and accessible
fraction for bosons, compared to the $\nu = 1/3$ found in the electronic case.
Choosing a lattice of  dimensions $L_x$ and $L_y$, and the number of
excitations, $N_p$,
fixes $\alpha = {2N_p}/{L_xL_y}$. As the  state space grows quickly, we
are constrained to the small systems in Table \ref{tab:overlap}.
	\begin{table}

	\begin{tabular}{c|ccc|ccc} \hline\hline  &
	   $L_x\times{}L_y$ & $N_p$ & Dim($H$) & $\alpha$ &
	   Laughlin & Transition($\Delta_c$) \\ & & & & & overlap  & \\\hline
	   $i$ & $4\times{}4$ & 2 & 512 & 0.25 & 0.89  & NA \\ 
	   $ii$ &$5\times{}5$ & 2 & 1250 & 
	   0.16 & 0.99 &  -1.1 \\
	   $iii$ & $6\times{}6$ & 2 & 2592&  0.11 & 0.99  &
	   2.5\\ 
	   $iv$ &$4\times{}4$ & 3 & 5472 & 0.37 & 0.29 &  -9.1\\
	   $v$ & $5\times{}5$ & 
	   3 & 20850 & 0.24 & 0.98 & -3.8\\ 
	   $vi$ & $6\times{}6$ & 3 & 62232 & 0.17 & 0.99 
	   & NA \\ \hline \end{tabular} \caption{\label{tab:overlap}
	   Results for systems of size $L_x\times{}L_y$ sites with $N_p$
	   particles. All systems have $C_1=1/2$ below the transition strength
	   $\Delta=\Delta_c$. Also shown is the Hilbert space dimensions
	   Dim$(H)$, and the Laughlin overlap.}\vspace{-.3cm}
	\end{table}

The simplest FQHE states are described by the Laughlin ansatz, which is an
exact solution for particles in a magnetic field with a contact interaction, and $\nu=1/q$.  Thus for a lattice, where the inter particle
interaction is only on site, such wavefunctions can be very good
approximations to the true groundstate.  The excitations of these states have abelian anyonic
statistics.
\begin{figure}

\includegraphics[width=8cm,height=8cm]{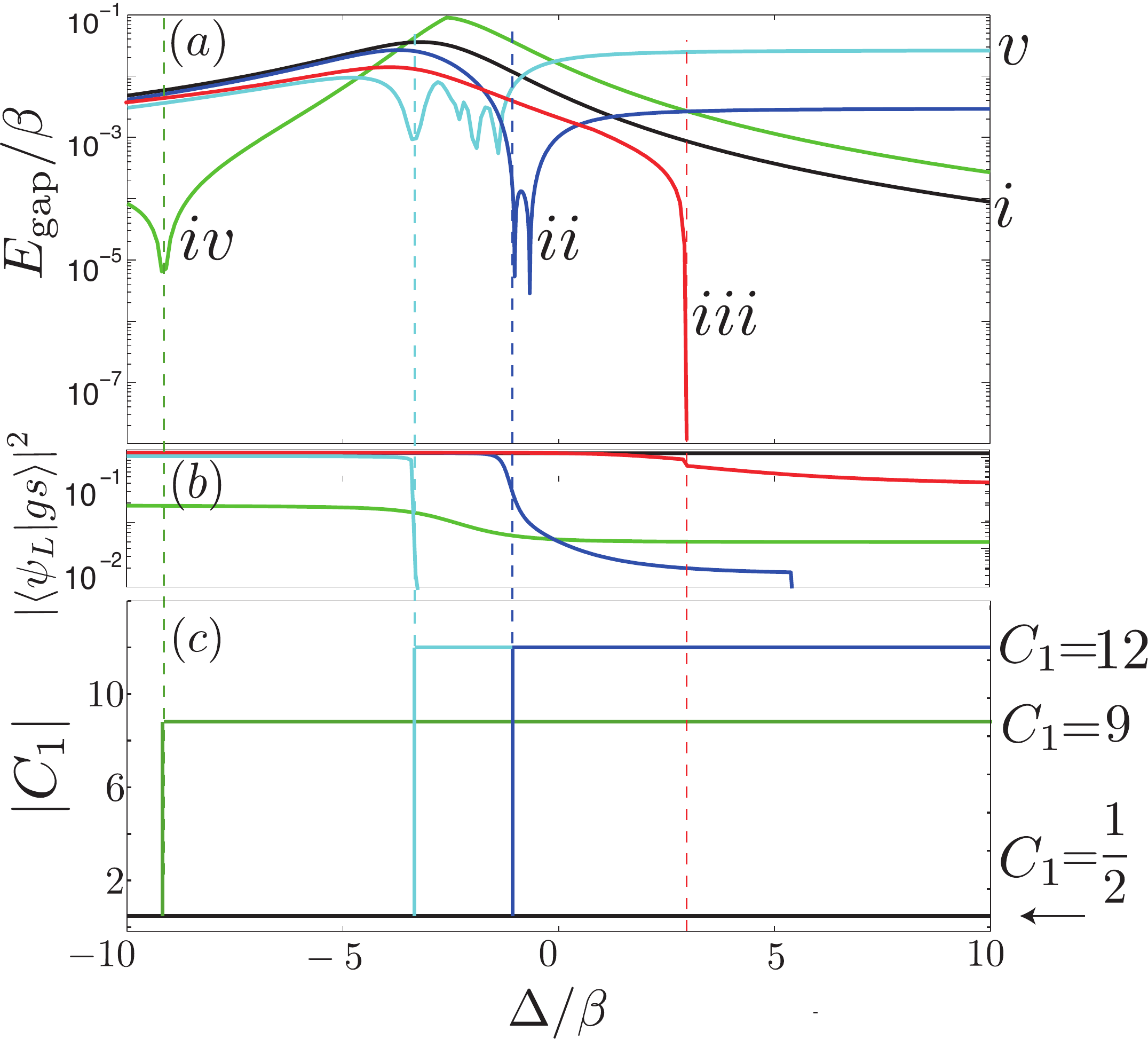}

\caption{\label{fig:gap}(a) Energy gaps, (b) Laughlin wavefunction overlap,
and (c) Chern numbers in the JCH FQHE as a function of the
detuning $\Delta$. For $\Delta << 0$, the effective on-site energy is very
high. In the opposite case, where
$\Delta >> 0$, the effective on-site energy is much lower than the energy from
the  pressure and the gap is due to multiple site occupancy. Dashed lines
indicate the transition from a fractional ${1}/{2}$ state to a
non-interacting particle state as determined by the Chern number. Colors
correspond to configurations in Table
\ref{tab:overlap}: $i$-black, $ii$-dark blue, $iii$-red, $iv$-green, $v$-light
blue.}\vspace{-0.3cm}
\end{figure}
The Laughlin wavefunction with periodic boundary conditions is:
\begin{equation}
\Psi_L(\bar{z}) \propto
F_{cm}(Z)f_{rel}(\bar{z})\prod\limits_{i}\psi^{L}_i,
\end{equation}
where $F_{cm}(Z)$ is a function of the centre of mass $Z
=\sum^{N_p}_iz_i$, and $f_{rel}(\bar{z})$ depends only on the relative particle
separations:
\begin{equation} \begin{array}{rcl} F_{cm}(Z) & = & \theta{}\left[
   {\begin{array}{c} N_p/q +(N_\phi{} -2)/2q \\ -(N_\phi -2)/q \\ \end{array}}
      \right]\left(q\frac{Z}{L_x}|iq\frac{L_y}{L_x}\right), \\ f_{rel}(\bar{z}) & = &
      \prod\limits_{i<j}^{N_p}\theta{}_1\left(\frac{z_i
      -z_j}{L_x}|i\frac{L_y}{L_x}\right)^q,
   \end{array} \end{equation} 
   where $\theta$ is the generalized Jacobi theta
   function\cite{Read1996}, and $\theta_1$ the odd theta function. $\psi_i^L$ are the single particle states $e^{-y_i^2/4}$ of the LLL.

For bosons (fermions) $q$ must be even (odd) so that $\Psi$ has the appropriate
 symmetry.  We define a version of the Laughlin wavefunction for states on a
 JCH lattice by replacing the single particle Landau wavefunctions
 $\psi^L$ with a corresponding  single polariton JCH wavefunction. 
This approximation allows the computation of the overlap between the explicit
groundstate of $H^{JCH}(\alpha)$ and the Laughlin ansatz. Direct product
states of single JCH states are not defined at points with multiple
exciations, however, as $\Psi(\bar{z}) = 0$ for any $z_i =z_j$
the issue is avoided.

Significant overlap between our numerical results and the Laughlin ansatz
can be a good indication of the FQHE. In Fig. 3(b) the overlap is shown as a
function of $\Delta$ for each configuration. In the hardcore boson
limit ($\Delta \gg 0$), our results match those found in \cite{Hafezi2007}. As
the onsite interaction
is reduced, multiple site occupancy can occur, and overlaps with the trial
wavefunction decreases. To fully quantify the groundstate, two additional
quantities need to be computed: (i) the topology of the state, and (ii) the
energy gap.
 The Chern number quantifies the topology and provides an unambiguous
indication of FQHE physics\cite{Tao86}:
\begin{equation}
C_1 =
\frac{1}{2\pi}\int{}d\theta_x\theta_y\left\langle{}\frac{\partial\Psi}{\partial\theta_x}\mid\frac{\partial\Psi}{\partial\theta_y}\right\rangle
-
\left\langle{}\frac{\partial\Psi}{\partial\theta_y}\mid\frac{\partial\Psi}{\partial\theta_x}\right\rangle,
\label{equ:chern}
\end{equation}
where $\theta_{x,y}$ are the generalized periodic boundary conditions on the
lattice:
\begin{equation}
   t^i_x(L_x)\Psi =  
   \Psi{}e^{i\theta_x} \qquad
t^i_y(L_y)\Psi  = 
\Psi{}e^{i\theta_y}
   \label{equ:GPBC}
\end{equation}
with $t^i_{x,y}$ the $x$ and $y$  magnetic translation operators on particle
$i$ . Varying $\theta_{x,y}$ induces
an electric field on the surface of the lattice, leading to the relationship
between the Hall conductance and Chern number via the Kubo formula: $\sigma_{H}
\propto C_1$ (see \cite{Tao86} for a full discussion). Hence, 
the topology of the groundstate with periodic boundary conditions  is directly
related to the quantization of the Hall conductance in the edged  geometry.
This measure provides a means of classifying the groundstate when the Laughlin
ansatz is no longer a good representation.  This occurs when lattice effects become significant, such as with 
large $\alpha$, when the single particle states deviate significantly from
the continuum LLLs.  Hafezi \emph{et al.}\cite{Hafezi2007} have found Chern numbers
for states in the BH model.

The degenerate groundstates, $\Psi_0(\theta)$, define a principal fiber bundle
over the $T^2$ manifold. The Chern number classifies the homotopy class of the
fiber bundle, which is a topological invariant.  That is, the Chern number is
insensitive to small perturbations relative to the energy gap. Only if the gap
closes can the transition to topologically different states occur.  Explicit
computation of Eq.(\ref{equ:chern}) is computational intensive. We instead use
the method first proposed in \cite{Hatsugai2004}, and used in
\cite{Hafezi2007}, which allows for efficient computation of the Chern number
in the presence of degeneracies. In this method, a phase is defined for
the ground state at each $\theta_{x,y}$ with respect to two reference states. The
Chern number is given as the signed sum of the vortices which occur at the
zeros of the overlap with one of the reference states.

Figures 3(a) and (c) plot the energy gap and Chern number respectivly as a
function of $\Delta$. Both
indicate that for some lattice configurations a transition occurs from the
FQHE state  to some uncorrelated
states.  In the JCH model (and BH), the discrete lattice gives
rise to pressure in the system. Competition between this pressure and the
on-site repulsion leads to a topological phase transition. In the limit of
weak interactions the pressure dominates, and the groundstate is defined by
single particle behaviour. Between these two limits the energy gap closes at a
single point in momentum space, marking the transition to a fractional state.
Shown in Table \ref{tab:overlap} for a range of
lattice configurations ($i-vi$) is the Laughlin wavefunction overlap and the
location of  the
value of $\Delta$ at which the transition occurs. We find that in the case of
configurations $i$ and $vi$, no such transition occurs due to an exact degeneracy in
the single particle energy spectrum. 
Before the transition, the ground state Chern number is $1/2$. The Chern
number changes discretely when the gap closes, to the non-interacting state.
The Chern number for the system is then the sum of Chern numbers for  single
particles, given by solutions to the Diophantine equation $C_1 =
sq/p-1/q$, $\{s, C_1\} \in \mathbb{N}$\cite{Goldman2009}. We find this to be the case for configurations
$ii,vii$ and $v$. For $iii$,  $C_1 $ is undefined due to the presence of  level
crossings, as we have also observed in the case of the BH model.  The
question of whether these transitions persist in the thermodynamic limit is
still an open question\cite{Hafezi2007}.

The proposed system can potentially be implemented in any cavity QED
framework. However, circuit QED systems are the most  promising as
they exhibit the largest atom-photon interactions, relative to cavity Q.
With
coherence times for qubits approaching 10$\mathrm{\mu{}s}$\cite{Paik2011}, and coupling strengths
$\approx 10^2\mathrm{MHz}$\cite{Bishop2009}, FQHE states on small lattices, on
the order of 15 sites,
could be produced. 

We have shown that many of the phenomena associated with the FQHE can be
realistically  emulated in cavity QED systems.
Cavity lattices allow direct inspection of quantum states, which offers an
unprecedented window into the physics of the QHE and topological phases.
The cavity QED framework also allows for very broad control over the system's
parameters and is readily extensible to more complicated configurations.
 For example, by including a three level atom with evenly
spaced levels, the photons can be given an effective 3-body contact interaction, for
which the well studied Pfaffian states are solutions\cite{Greiter1992}. These states have
non-abelian statistics, and are a basis for the description of the $\nu =
{5}/{2}$ Quantum Hall plateau.
An implementation of the system considered here is an
exciting prospect for the near future and will provide crucial insight into
the physics of the Quantum Hall Effect.

 A.D.G.~acknowledges the
Australian Research Council for financial support (Project
No.~DP0880466).

\bibliographystyle{apsrev}
\bibliography{qhe}

\end{document}